\documentclass[twocolumn,showpacs,preprintnumbers,amsmath,amssymb,prl,floatfix]{revtex4}
\usepackage{graphicx}
\usepackage{dcolumn}
\usepackage{bm}

\begin{document}

\title{Recombination fluorescence in ultracold neutral plasmas}

\author{S. D. Bergeson}
 \affiliation{Department of Physics and Astronomy, Brigham Young
 University, Provo, UT 84602, USA}

\author{F. Robicheaux}
 \affiliation{Department of Physics, Auburn University,
 Auburn, Alabama 36849, USA}

\date{\today}

\begin{abstract}
We present the first measurements and simulations of recombination
fluorescence in ultracold neutral plasmas.  In contrast with
previous work, experiment and simulation are in significant
disagreement. Comparison with a recombination model suggests that
the disagreement could be due to the high energy portion of the
electron energy distribution or to large energy changes in
electron/Rydberg scattering. Recombination fluorescence opens a new
diagnostic window in ultracold plasmas because it probes the
deeply-bound Rydberg levels, which depend critically on electron
energetics.
\end{abstract}


\maketitle

Ultracold neutral plasmas can provide new insights into relaxation
phenomena in strongly coupled Coulomb systems.  They are created by
photoionizing laser cooled atoms and therefore have very precisely
controlled initial temperatures and densities \cite{killian99,
simien04, cummings05a, killian07}. Photoionization causes an
impulsive hardening of the interparticle potential \cite{murillo06},
and the pathway to (non)equilibrium can be studied in detail
\cite{pohl05c, pohl06}.

The electron system equilibrates on the shortest time scales and
largely determines how the plasma expands \cite{ kuzmin02a,
robicheaux02, mazevet02, pohl04a, fletcher06}.  Indirect
measurements of the electron temperature agree well with simulations
\cite{robicheaux02, roberts04, gupta07}. Disorder-induced heating,
three body recombination (TBR), and electron/Rydberg scattering are
important heating mechanisms in the plasma.  The latter two are
thought to occur on times that are long compared to the electron
plasma frequency. Thus TBR can probe the electron system at early
times in the plasma evolution after disorder-induced heating has
occurred \cite{fletcher07}.

In higher temperature plasmas, the TBR picture is an electron and
ion colliding in the presence of another electron.  The TRB rate is
the two-body collision frequency ($n \sigma v_{th}$) multiplied by
the probability that a third body is one collision distance away ($n
b^3$).  The collision distance and cross section depend on the
electron temperature ($b=e^2/4 \pi \epsilon_0 k_B T$ and $\sigma=
\pi b^2$), giving the well-known TBR rate $\alpha_3 = {\cal R} n^2
T^{-9/2}$, where ${\cal R}$ is the TBR rate coefficient
\cite{formulary}.

In a strongly-coupled system, this binary collision picture breaks
down.  When the nearest-neighbor potential energy exceeds the
kinetic energy ($\Gamma \equiv e^2 n^{1/3}/4 \pi\epsilon_0 k_B T >
1$), the plasma is strongly coupled. The average distance between
electrons becomes comparable to the collision distance $b \sim
n^{-1/3}$.  The electrons are in a constant collision and TBR
becomes, in a sense, many-body recombination. A numerical simulation
of early-time recombination suggest that even moderate initial
coupling in the electron system changes the recombination rate by
perhaps a factor of two \cite{kuzmin02b}. Other theoretical
treatments suggest larger changes \cite{hahn97,hahn00}. However,
there is no experimental evidence that TBR departs from the standard
formulas \cite{comment1}.

In this paper we present a new study of three-body recombination in
ultracold neutral plasmas.  We measure the time-dependent
fluorescence emitted by the plasma following electron/ion
recombination for a range of initial plasma densities and electron
temperatures.  The fluorescence signal is the end result TBR,
electron-Rydberg collision, and radiative cascade. Thus, it is
sensitive to the details of the evolving electron system.  We
compare these results with a numerical simulation and find
significant disagreement.  A recombination model suggests that this
is related to the high-energy portion of the electron energy
distribution or to large energy changes in electron/Rydberg
scattering.



Our ultracold plasmas are created by photoionizing laser-cooled
calcium atoms in a magneto-optical trap \cite{cummings05a,
cummings05b}. The plasma density profile is approximately Gaussian
and spherically symmetric with the density $n(r,t)=n_0
\exp(-r^2/w^2(t))$ and $w(0)=0.5$ mm. The initial ion temperature is
equal to the neutral atom temperature ($\sim$1 mK). The initial
electron energy is equal to the difference between the
photoionization laser photon energy and the atomic ionization
energy.  The lower limit is set by the laser bandwidth and by
continuum lowering effects to approximately $2E_e/3k_B = 0.5$K.  Our
two-step photo-ionization process ionizes nearly 100\% of the atoms
in the MOT, and our maximum plasma density is set by the MOT density
to $n \leq 2\times 10^{10}$ cm$^{-3}$.

\begin{figure}
\begin{center}
\includegraphics[width=3.4in]{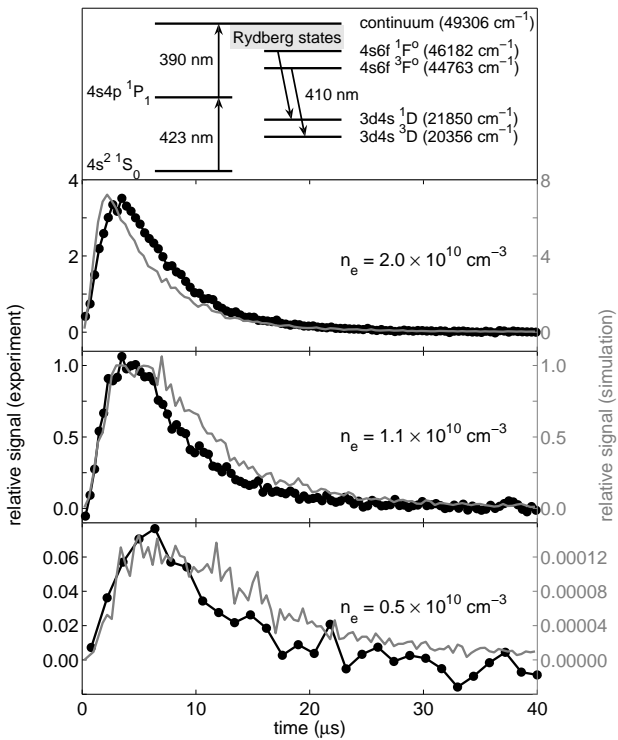}
\end{center}
\caption{Top panel: Partial energy level diagram for neutral
calcium.  For a fraction of recombination events, the radiative
cascade produces fluorescence at 410 nm. Lower panels: Recombination
fluorescence signal at 410 nm for $T=57$K and three different
densities, experiment (black points, left scale) and simulation
(gray line, right scale). The simulation changes more much rapidly
with density than the experiment. \label{fig:level}}
\end{figure}

Fluorescence from the plasma is collected by a fast lens system and
detected using a PMT. An interference filter blocks light from the
lasers and the MOT. Most of the fluorescence results from emission
at 410 nm.  A typical fluorescence signal and partial energy level
diagram are shown in Fig. \ref{fig:level}.  At early times, the
fluorescence signal grows as recombination proceeds.  After a few
$\mu$s the plasma expands and the density decreases. The
fluorescence signal falls as recombination slows down and the plasma
moves out of the view of the fluorescence collection optics.

The fluorescence signal is the end product of a series of collision
events \cite{robicheaux04}.  Free plasma electrons and recombined
Rydberg atoms are in a quasi-equilibrium. Collisions form and then
ionize Rydberg atoms, most of which have a binding energy roughly
equal to $k_B T$. This collisional equilibrium gives rise the
so-called ``thermal bottleneck,'' which is a minimum in the Rydberg
state distribution near $-4k_B T$ \cite{kuzmin02b}. A collision
occasionally occurs that puts the atom in a Rydberg state below the
bottleneck.  Additional collisions and radiative cascades move a
fraction of these atoms into the fluorescing state.



We have performed a numerical simulation of recombination and
fluorescence in the plasma using a method similar to Refs.
\cite{robicheaux02, robicheaux03}. Like the experiment, we assume
the plasma to be nearly charge neutral with a spherically symmetric
Gaussian density distribution.  The electrons are in thermal
equilibrium at a temperature $T$. The plasma expands self-similarly,
and the density can be written as $n(r,t)=n_0 \exp[-r^2/w^2(t)]$.
The equations of motion at this level of approximation are described
in Sec.~IIIc of Ref.~\cite{robicheaux03}. The main trends are that
the expansion decreases the thermal energy of the electrons while
increasing the radial speed of the ions. When recombination occurs,
the density decreases slightly and the temperature increases. An
electron collision with an atom that decreases the atom's energy
increases the electron temperature and vice versa.

We compute the evolution of the atom population using a Monte Carlo
technique. During a time step, we compute the probability of
formation of an atom at each position $r$ and which state it is in
using the method of Sec.~IV of Ref.~\cite{robicheaux03}. The method
gives more atoms formed at the center of the plasma where the
density is high and the atoms are mostly formed in high Rydberg
states \cite{pajek97, pajek99}.
 The radial velocity of the atom is taken to be
the radial velocity of an ion at the position the atom is formed;
after the atom is formed, we track its position and velocity so we
know the correct electron density near each atom and we know whether
the atom is within the detection region when the photon is emitted.
After the atom is formed, its state changes due to electron
collisions and radiative decay. The electron collisions can reionize
the atom, excite it to higher energies, or de-excite it. The
electron collision processes are computed with a Monte Carlo
technique using the rates of Ref.~\cite{mansbach69}. One difference
with Ref.~\cite{robicheaux03} is that we used the exact radiative
decay rates for $n\ell\rightarrow n'\ell'$ for $n\leq 30$. We did
this because the $\ell$ states are not necessarily statistically
mixed for lower $n$ and the fluorescence signal depends on getting
the cascade correct. When an atom's principle quantum number drops
below 5 it is counted as a fluorescence photon and removed from the
simulation. The calculation has only one free parameter related to
the location of the thermal bottleneck. The formulas in Refs.
\cite{pajek97, pajek99} lead to a bottleneck at $-2k_B T$.  Changing
this value changes the shape of the fluorescence curves somewhat
because the details of the fluorescence signal depend on the high
energy portion of the electron distribution. However, there was no
choice for the bottleneck that allowed us to reproduce all of the
measurements.



The simulation and experiment are compared in the lower three panels
of Fig. \ref{fig:level}.  In these data, the initial electron
temperature is $T=2E_e/3 k_B = 57K$, and the initial plasma size is
$w(0) = 0.5$ mm.  Qualitatively, the simulation and experiment are
in reasonable agreement.  For the lower two densities, the rising
edge of the simulation and experiment have the same time dependence.
However, there are quantitative mismatches.  The relative change in
signal strength with density is completely different for the
simulation and experiment.  In addition, the time dependence of the
high density simulation does not match the experiment.

This disagreement is surprising.  Previous work has demonstrated
good agreement between this simulation and the experiment.  This
simulation and related approaches give the proper plasma expansion
velocity, electron temperature, and even the high $n$ Rydberg state
distribution. However, recombination fluorescence opens a new window
in these ultracold plasmas because it probes the deeply-bound
Rydberg levels.  The population in these levels depends critically
on the high-energy portion of the electron distribution.



Some insight can be gained by comparing to a simple rate equation
model. The rate at which Rydberg atoms are formed in the plasma can
be written as

\begin{equation}
\frac{dn_r}{dt} = -{\cal S} n_r n_e + {\cal R} n_e^3 T_e^{-9/2},
\label{eqn:dnrdt}
\end{equation}

\noindent where ${\cal S}$ is the collisioinal ionization rate,
${\cal R}$ is the three-body recombination rate coefficient, $n_r \;
(n_e)$ is the Rydberg atom (electron) density, and $T$ is the
electron temperature.  This simplified model treats all Rydberg
levels as essentially equal and neglects transitions between
different levels. It neglects radiative decay and assumes the plasma
is charge neutral. Recombination increases the electron temperature,
and the corresponding rate equation can be written as

\begin{equation}
\frac{dT_e}{dt} = \frac{1}{n_e}\frac{dn_r}{dt} \left[ T_e + T_0
\Gamma_0 \left(\frac{n_e}{n_e(0)}\right)^{1/3}\right].
\label{eqn:dTdt}
\end{equation}

\noindent where $T_0$, $\Gamma_0$, and $n_e(0)$ are the initial
electron temperature, strong coupling parameter, and electron
density.  The total energy gained in one recombination event is the
sum of the colliding electron energy and the correlation energy. We
use the relation $n_r + n_e = n_e(0)$ to write these coupled
differential equations for $n_r(t)$ and $T_e(t)$ in terms of known
parameters \cite{modelNote}.

The relationship between the Rydberg atom density and the
fluorescence emission rate is non-trivial.  However, there is some
evidence from our simulation that a single collision de-excites the
Rydberg atom to radiating states. This suggests that the Rydberg
atom density at a time $t_1$ is proportional to the fluorescence
signal $s(t)$ integrated to a time $t_2$,

\begin{equation}
n_r(t_1) \sim \int_0^{t_2} s(t) dt. \label{eqn:nr}
\end{equation}

\begin{figure}[t]
\begin{center}
 \includegraphics[width=3.5in]{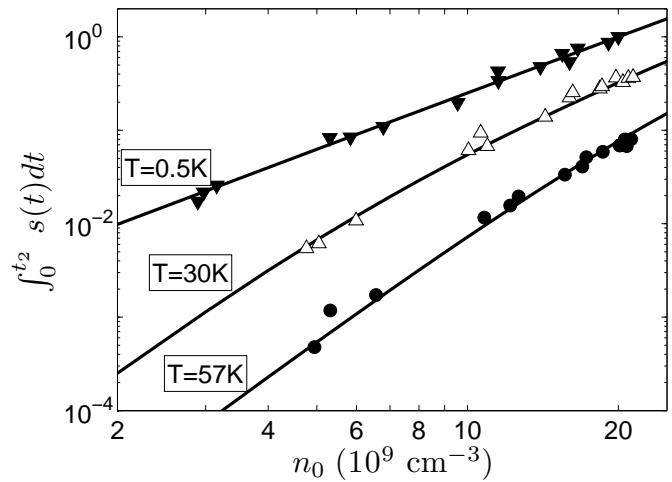}
 \end{center}
\caption{A plot of Eq. \ref{eqn:nr} comparing the rate equation
model (lines) to the experiment (points) for different temperatures
and densities using $t_1 = t_2 = 1.5 \;\mu$s and a single overall
proportionality constant.
 \label{fig:out1}}
\end{figure}

\noindent A comparison of this simplified rate equation model to the
experiment is shown in Fig. \ref{fig:out1}.  For this comparison, we
choose $t_2$ early enough that the plasma has not started to expand.

The relatively good agreement between the rate equation model and
the data and the poorer agreement of the simulation deserve some
attention.  The simulation is much more complete than the model.  It
uses the most widely accepted approach for Rydberg atom formation,
for Rydberg/electron collisions, and for Rydberg state distribution.
It uses the exact hydrogenic decay rates.  It also includes
geometrical effects of the fluorescence collection optics.  On the
other hand, the rate equation model doesn't include any effects due
to Rydberg state distribution or to the cascade down into
fluorescing states.  Perhaps, then, the simulation tells us
something about effects beyond the ``standard'' cold plasma model.
It is likely that these are related to details of the high-energy
portion of the electron distribution.  A non-Maxwellian tail would
change the location of the thermal bottleneck, the deeply-bound
Rydberg state distribution, and the cascade rates.



A simulation in Ref. \cite{kuzmin02b} suggested that strong coupling
in the initial electron system might change the three-body
recombination rate by perhaps a factor of two.  Guided by the rate
equation model, we can explore this possibility.  For early enough
times, the collisional ionization term, ${\cal S}$, in Eq.
\ref{eqn:dnrdt} can be neglected because the recombined Rydberg atom
density is small.  With this assumption, the Rydberg atom density is
$n_r(t)/n_e(0) \sim n_e^2(0)T^{-9/2}t$.  Using Eq. \ref{eqn:nr} and
the definition of the strong coupling parameter, we can write

\begin{equation}
\Gamma \sim \left\{ [n_e(0)]^{-3/2} \int_0^{t_2} s(t) dt
\right\}^{2/9} \label{eqn:g1}
\end{equation}

\noindent For the low density, high temperature data, the strong
coupling parameter can be reliably calculated from the initial
experimental values \cite{gupta07}.  However, when the plasma is
generated closer to the strongly-coupled regime, the electron
temperature rapidly increases and the density decreases to values
that can be significantly different from the initial ones.  However,
it is expected that the recombination rates are smooth functions of
the strong coupling parameter.

\begin{figure}
\includegraphics[width=3.4in]{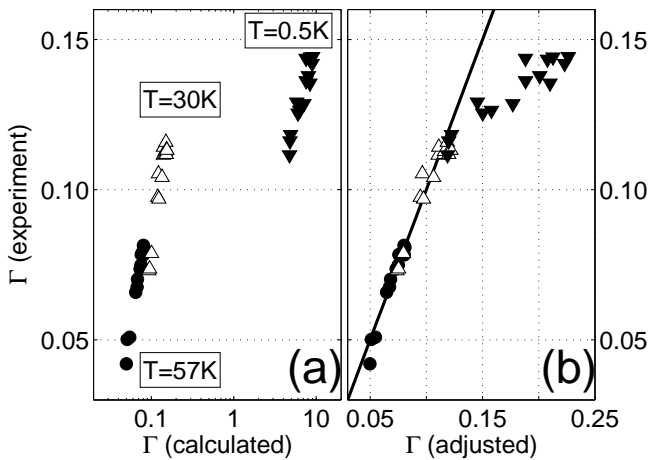}
\caption{Strong coupling parameter $\Gamma$ from Eq. \ref{eqn:g1}
plotted vs. (a) initial calculated $\Gamma$ and (b) the
temperature-adjusted $\Gamma$.  The experimental data is taken from
Fig. \ref{fig:out1}.  In panel (b) the adjusted temperatures are
adjusted as follows: Black circles ($T=57$K) no adjustment, upright
triangles ($T=30$K) adjusted to 38K, inverted triangles ($T=0.5$K)
adjusted to 20K. The solid line plots the equality $\Gamma \;
\mbox{(experiment)} = \Gamma \; \mbox{(adjusted)}$.
\label{fig:gamma}}
\end{figure}

In Fig. \ref{fig:gamma}(a) we plot the strong coupling parameter
extracted from the experiment vs. the calculated value.  The
experimental data is scaled to agree with the calculated value for
the smallest $\Gamma$.  The higher $\Gamma$ data are in obvious
disagreement. However, by adjusting the electron temperature in the
calculation, the experimental $\Gamma$ can be made to vary smoothly
as shown in Fig \ref{fig:gamma}(b). Beginning at $\Gamma\approx 0.1$
the experimental $\Gamma$ falls well below the adjusted value.  This
is what one would expect if the three-body recombination rate is
depressed due to many body interactions.  Of course, this is also
what one would expect if the electron temperature is simply higher
than expected.  An independent and accurate electron temperature
measurement at early times is obviously needed, although if the tail
of the electron distribution is non-Maxwellian, the meaning of the
temperature measurement becomes less clear.

In conclusion, we have demonstrated a new fluorescence technique to
study three-body recombination in ultracold neutral plasmas.  This
method probes the evolution of deeply-bound Rydberg states.  In
contrast with previous work, where the evolution of plasma
parameters can be predicted using simple fluid models, we find
substantial disagreements. This work suggests that the high energy
electron distribution or the Rydberg state distribution is different
from that predicted by the standard cold plasma simulation.  We note
that when the ultracold plasmas start with the electron strong
coupling parameter in the range of $0.1<\Gamma<10$, the early time
evolution of the electron and Rydberg state distributions are not
well known.  Measuring the plasma fluorescence provides an
opportunity to study this regime.

This work is supported in part by Brigham Young University, the
Research Corporation, the National Science Foundation (Grant No.
PHY-0601699) and the Chemical Sciences, Geosciences, and
Biosciences Division of the Office of Basic Energy Sciences,
U.S. Department of Energy.


\begin{thebibliography}{27}
\bibitem[{com()}]{comment1}
\bibinfo{note}{Early experimental publications suggested that the recombination
  rate was in disagrement with plasma theory, however later simulations that
  included both recombination and scattering brought theory and experiment into
  agreement.}
\end{thebibliography}
\end{document}